# Controlling photon emission from silicon for photonic applications

Seref Kalem

TUBITAK-BILGEM National Research Institute of Electronics and Cryptology, Gebze 41470 Kocaeli, Turkey

## ABSTRACT

The importance of a photon source that would be compatible with silicon circuitry is crucial for data communication networks. A photon source with energies ranging from UV to near infrared can be activated in Si as originationg from defects related to dislocations, vacancies, strain induced band edge transitions and quantum confinement effects. Using an etching method developed in this work, one can also enhance selectively the UV-VIS, band edge emission and emissions at telecom wavelengths, which are tunable depending on surface treatment. Deuterium $D_2O$ etching favors near infrared emission with a characteristic single peak at 1320 nm at room temperature. The result offers an exciting solution to advanced microelectronics The method involves the treatment of Si surface by deuterium Deuterium containing acid vapor, resulting in a layer that emits at 1320 nm. Etching without deuterium, a strong band edge emission can be induced at 1150 nm or an emission at 1550 nm can be created depending on the engineered surface structure of silicon. Schottky diodes fabricated on treated surfaces exhibit a strong rectifying characteristics in both cases.

## 1. INTRODUCTION

The need of semiconductor industry for silicon based light source working at optical communication window is becoming increasingly important for advanced technology nodes. The light source that is operational at one of these wavelengths is a key parameter in a number of applications ranging from interconnects in system on chip, multi-chip modules to printed circuit boards and free space optical communications[1]. As an indirect band gap material, silicon cannot emit a useful light at room temperature for applications. However, a number of reports have been made on defect related luminescence in Si. Light emission can usually be observed in near infrared (NIR), which are mainly due to dislocations. But, the intensity of these emissions is relatively weak at room temperature and they are only originating from the locations where the dislocations are found. Demonstrated devices have a broad and multiple peak emission bands [2]. The NIR light sources obtained by the approach described here can form a basis for the realization of new Si based optoelectronic and photonic devices. For their ability to pass through layers commonly used in Si IC's without getting absorbed, new optical transistors with light sources have been investigated by various methods and device structures [3]. It is also possible to observe light emission in the infrared region (800-1800 nm) from silicon containing Silicon and Germanium quantum dots. Moreover, it has been demonstrated that the down scaling and strain enhance the band edge emission at around 1100 nm. On the other hand, point defects and impurities can be the potential sources leading to sub-band gap emissions [4].

## 2. EXPERIMENTAL

In order to produce a light emitting Si and related device at NIR, the surface of silicon wafer is treated by hydrogen(H2O) or heavy water or deuterium oxide $D_2O$ containing acid vapor. A chemical mixture of $HF:HNO_3:H_2O$ and $HF:HNO_3:D_2O$ was used to form the acid vapor, where the weight ratio of HF solution can be between 40% and 50% and that of $HNO_3$ solution is between 60% and 70%. The treatment is carried out at room temperature in a teflon chamber. The weight ratios of HF and $HNO_3$ used in our experiments were typically 48% and 65%, respectively. The solutions are of semiconductor applications grade or GR for analysis. $HF:HNO_3:D_2O$ mixture can be formed using 2 to 7 unit volume from HF, 1 to 12 unit volume from $HNO_3$ and 1 to 6 unit volume from $D_2O$. As wafers, Silicon with p-type (5-20 Ohm-cm) and n-type (5-10 Ohm-cm) conductivities, <100> and <111> crystallografic orientations were used for treatment. Si surfaces treated by hydrogen or deuterium were rinsed in DI water to remove the $(NH_4)_2SiF_6$ (hereafter ASH) and $(ND_4)_2SiF_6$ (hereafter DASH) layers. The photoluminescence was excited by an Ar ion laser and the emission was collected by 0.5m imaging grating monochromator equipped with a liquid nitrogen cooled InGaAs arrayed detector.

## 3. RESULTS and DISCUSSION

There is no work demonstrating a single peak emission at 1320nm from silicon. The strong photon emission generated by treating the wafer using heavy water containing acid vapor ($HF:HNO_3:D_2O$). Furthermore, no electronic device has been produced using the same method and technology. Deuterium has been used for different purposes in semiconductor research and manufacturing. Deuterium was used in the production of low-loss optical materials for optical communication. In another example, the improvement of MOS structures has been carried out by annealing in deuterium atmosphere. In the following example, deuterium was used with HF vapor for cleaning and passivation of Si wafer surface. It has been shown that deuterium is satisfying dangling bonds very effectively at Si-SiO2 interfaces and thus reducing trap density which leads to improved CMOS characteristics at low voltages [5]. Another work showed that deuterium was improving photoluminescence by reducing degradation[6]. It was shown that MOS transistors, which were fabricated on wafers annealed in deuterium atmosphere enhanced the performance of the transistors. In such examples, it was shown that the energy coupling of Si-D bonds to Si-O-Si TO mode and Si-O modes was dramatically enhanced in D treated $SiO_2$ [7].

The present work shows that the photon emission wavelength from Si can be tuned by changing etching conditions. For example, a new silicon (Si) semiconductor light source that emits a single peak at 1320 nm in the infrared between 800 nm and 1800 nm with a full width at half maximum of less than 200 nm and relatively high photoluminescence quantum efficiency at room temperature. The result offers an exciting solution to advanced Si fabrication technology by providing a photonic component to Si circuitry where silicon is known to be a poor light emitter due to its indirect band gap. The method involves a silicon surface treated by an acid vapor containing heavy water or deuterium oxide $D_2O$, resulting in a layer that emits at 1320 nm. The resulting surface condition following this treatment was investigated by luminescence, ellipsometry, infrared spectroscopy, scanning electron microscopy and electrical measurements. Based on this investigation, we find that the surface layer contains 24 at.% of oxygen, which is relatively high concentration level as compared to similar surface treatment methods. Electrical measurements indicate that $Au/Ti/SiO_x$ surface layer contact fabricated on this structure exhibits a Schottky diode chracteristics with a strong rectifying behavior.

A part of the background hereof lies in the development of the surface treatment work of Si wafers by acid vapor during which ammonium silicon hexafluoride $(NH_4)_2SiF_6$ is formed on silicon wafer surface [8]. The silicon light source emitting only a single peak at 1320 nm with more than 50% luminescence quantum efficiency was produced for the first time using deuterium oxide $D_2O$ containing $HF:HNO_3$ acid vapor. This work differs from others by a single emission peak in the infrared band from 800 nm to 1800 nm. In other words, this application is promising for a process of making two and three terminal devices such as sensor, laser, and light emitting devices (LED) using the new light emitting properties of silicon wafer.

### 3.1 Structural

Figure 1 shows the SEM image of the structure formed on the surface of a wafer when the wafer is treated by the vapor of $HF:HNO_3:D_2O$ chemical mixture. The detailed description of the experimental apparatus can be found elsewhere [8]. Following the treatment, a surface layer that is composed of a deuterated ammonium silicon hexafluoride layer $(ND_4)_2SiF_6$ (hereafter DASH) and a SiOx interface layer was formed. This layer contains also $(NH_4)_2SiF_6$ (hereafter ASH) due to the presence of $H_2O$ or hydrogen in the chemical mixture. When the deuterated ammonium silicon hexafluoride layer $(ND_4)_2SiF_6$ was removed from the surface by rinsing the wafer in DI water, a nanostructured surface remains as shown in Fig. 1. The nanostructured surface layer is composed of silicon nanocrystals, which are encapsulated by SiOx where x value is close to 2 as estimated from the Si-O-Si stretching mode position from FTIR analysis. At a closer view, it can readily be seen that this structure is composed of nanometer sized coalescent grains, which are connected to each other in chains. The chain-like structure is separated by large air gaps of up to about 100 nm. The size of the grains can be down to few nanometers. We also observe hillocks-like structures with sizes up to100 nanometer, which can ease the injection of current. It is possible to produce a nanostructured features on silicon surface by various methods. Nanostructured silicon has been studied by a number of groups but none of them has reported a single peak emission at 1320 nm with a high quantum efficiency [9][10]. Deuterium induces exciting changes in silicon wafer surface as evidenced from our investigation results.

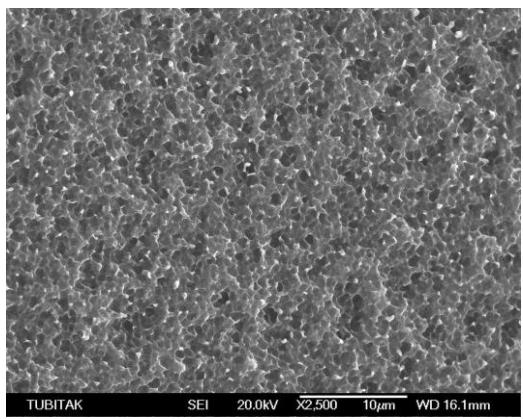

**Figure 1.** Surface morphology of the layer formed following the treatment of the wafer surface by the vapor of $HF:HNO_3:D_2O$ acid mixture. The layer containing the ASH and DASH was removed by rinsing the wafer in DI water. The EPS measurements indicate that the surface layer contains about 24 at. % Oxygen and 86 at. % Silicon.

## 3.2 Vibrational modes

FTIR investigation shows that deuterium interacts with silicon and forms bonds within the DASH matrix, which can be removed from the surface depending on the application by rinsing the wafer in deionized water. The presence of this layer and the role played by deuterium atom in forming the surface were determined by FTIR analysis. Figure 2 shows the FTIR spectra of the wafers treated by $D_2O$ containing acid vapor with p-Si(111) for 6 hours. FTIR investigation revealed the changes in microscopic structures of the Si wafer which was treated by heavy water incorporated acid vapor. The intensity of the oxygen band (Si-O-Si) observed at 1090 cm-1 is very strong at early stages of the treatment, proving that there is an intense oxidation process going on right at the beginning of the reaction between the silicon wafer and the deuterated acid vapor. The relative weakness of the Si-O-Si stretching modes at later stages of the process (6 hours of treatment) as shown in Figure-2 supports the stronger oxidation at the beginning of the process. Moreover, we find that the Deuterium plays an essential role in the formation of the new surface structure. FTIR investigation provides the evidence for the presence of bonded deuterium atoms within the ASH matrix. This was demonstrated by the presence of N-D vibrational modes. The vibrations which were observed in Figure-2 at 2524 cm$^{-1}$ and 2425 cm$^{-1}$ belong to N-D and N-$D_2$ vibrations. Because these modes were shifted by 1.37 with respect to N-H modes at 3132 cm-1. This number is indeed very close to the mass ratio of deuterium and Hydrogen atoms, that is $m_D/m_H$ = 1.376. Thus, the replacement of H atoms by D atoms shifts the frequency to 2425 cm$^{-1}$. We attribute the bands at 2425 cm$^{-1}$ and 1260 cm$^{-1}$ to stretching and bending vibrations of N-D modes and their presence and intensity are indicative of the presence of deuterium. FTIR reveals three Si-O stretching modes usually observed in SiOx at 1092 cm$^{-1}$, 1128 cm$^{-1}$ and 1183 cm$^{-1}$. These modes can be readily assigned based on previous reports as resulting from Si-O-Si stretching modes. In addition, a doublet with a stronger intensity appears at 1263 cm$^{-1}$ and 1272 cm$^{-1}$, which can be identified as the occurrence of two pairs of closely spaced N-D stretching doublets [11].

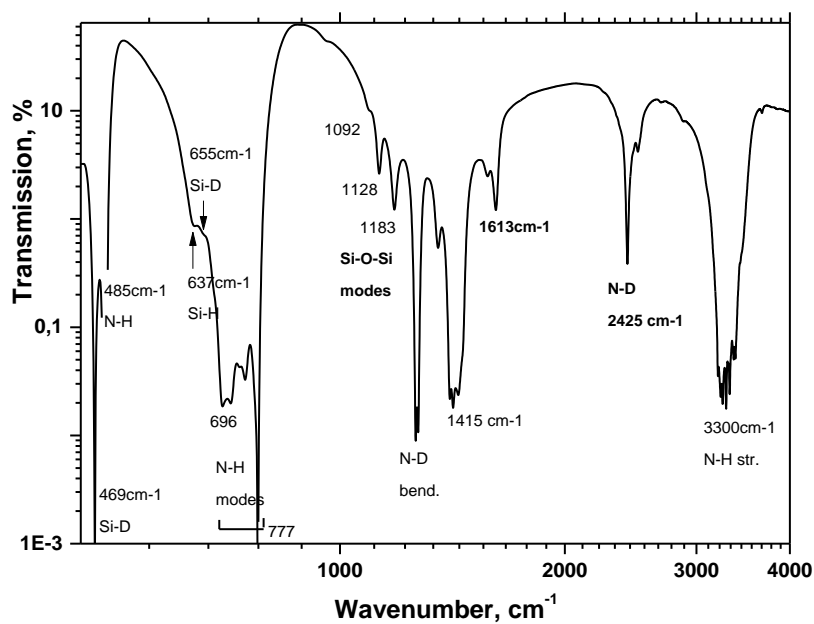

**Figure 2.** Infrared vibrational modes of N-D and N-H related bonds. Si-O-Si stretching modes are located at around 1100 cm-1 for a sample, which was treated for 6 hours by Deuterium containing vapor.

There is a very thin nanostructured silicon oxide (SiOx) layer formed on the surface of the Si wafer treated by $D_2O$ containing acid vapor, wherein **x** is around 1.8 as estimated from the peak position of Si-O bond [12]. Its thickness can range from few nanometers up to few hundred nanometers depending on the treatment parameters. SEM-EDS measurements indicate that the atomic concentration of oxygen within this layer can be as high as 25%. This result shows that nanostructured silicon is oxidized or grains on the surface are covered by an oxide layer.

The presence of Si nanostructures and its effects on optical properties can be understood from the data obtained by SE analysis. As deduced from the $2^{nd}$ derivative of the dielectric function $\varepsilon(\omega)$, where several critical energy points can be observed despite the rough surface. We attribute these critical points (CP) to $E_1$ and $E_2$ band energies of bulk Si and Si nano crystals. Energy points at 3.43 eV and 4.13 eV are likely due to E1 and E2 critical points in the bulk Si, respectively. E1 CP, the direct band gap edge, is broad and strongest, featuring the domination of direct transitions. Whereas the CP at 4.65 eV could be attributed to Si nano crystals, in which the quantum confinement is in effect. The region at around 700 nm where the interference fringes was wiped out is likely due to an effective bandgap of the Si nanocrystals. The presence of a PL at around 1.76 eV and at higher energies (around 2.0 eV) supports these findings. Peak groupings observed at around 900 nm could be associated with a band gap of larger nanocrystals. From the PL energy, one can estimate an effective crystal size ranging from 5 nm down to 1 nm based on previous studies.

**3.3 Controlling peak position**

It was found that any one of the NIR photon emission peak positions can be favored by controlling etching conditions. Silicon wafer surface when treated by a hydrogen containing vapor, a strong single peak at 1150nm region can be generated as shown in Fig. 3. When the wafer is treated by a Deuterium containing vapor, a strong single peak at 1320 nm telecom wavelength can be produced. This is a very significant change in the NIR region, inducing a peak at a desired telecom wavelengt. Detailed investigations shown that silicon wafer treated by $D_2O$ containing acid vapor exhibit exciting unique optical properties. Figure-3 shows a typical photoluminescence signal obtained from silicon treated with heavy water containing acid vapor. The spectrum indicates the presence of a peak at 1320 nm with a full width at half maximum of about 190 nm when the sample is excited with an Argon laser line of 488 nm at room temperature. This emission can be originated from D3 dislocations at misoriented bonding of wafers [13] and dislocation loops [14] or from impurities. However, their luminescence was observed at low temperature, which is accompanied by other emissions at near band edge at 1100 nm and D1 dislocations at 1500 nm. Actually, it is possible to deconvolute this band into two emission bands at 1306 nm (0.949 eV) and 1391 nm (0.891 eV) with a separation of 58 meV, thus suggesting TO phonon replicas. If our emission is related to D3, then our method is very effective in activating this process. The experimental demonstration of the presence of a silicon having a single peak emission at 1320 nm between 800-1800 nm with more than 50% photoluminescence quantum efficiency is very important for applications. Thanks to this finding, one can provide the semiconductor industry with the most needed silicon light source enabling photonic functions. There is no similarity between this result and other silicon light sources having photon emission in the infrared. This is the unique silicon light source in infrared region with more than 50% quantum efficiency and these features have not been previously observed. For the efficiency estimation, it was assumed that 50% of the laser power was transmitted through the sample and about 25% is reflected. The peak position of the NIR photopn emission can be shifted to 1550nm telecom wavelength, by inducing a rougher surface structure.

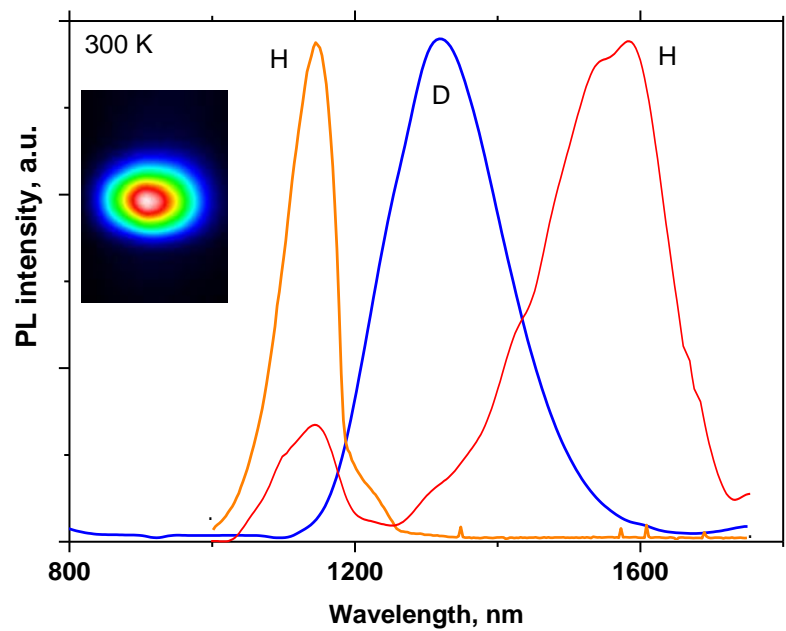

**Figure 3.** Room temperature photoluminescence from the p-silicon wafer that was treated by H and D containing acid vapor. HF:HNO$_3$:H$_2$O treatment results in a strong band-edge emission while HF:HNO$_3$:D$_2$O acid mixture induces a strong photon emission at 1320nm. The insert is the image of the visible photoluminescence originating from a photonic structure consisting of Si rods of 500 nm in diameter.

Figure 4 is a sketch of the band structure indicating possible radiative recombination pathways inducing NIR photon emission at telecom wavelengths on Si wafer, which is treated by the said method. It is very likely that the 1320 nm photon source is due to defects involving dislocation lines. However, a possible involvement of the defect states at interfaces can not be ruled out from this very efficient light generating process. The quantum confinement nature of the surface structure enhances the number of electron-hole pairs with an overlapping wavefunctions thus leading to high quantum efficiency.

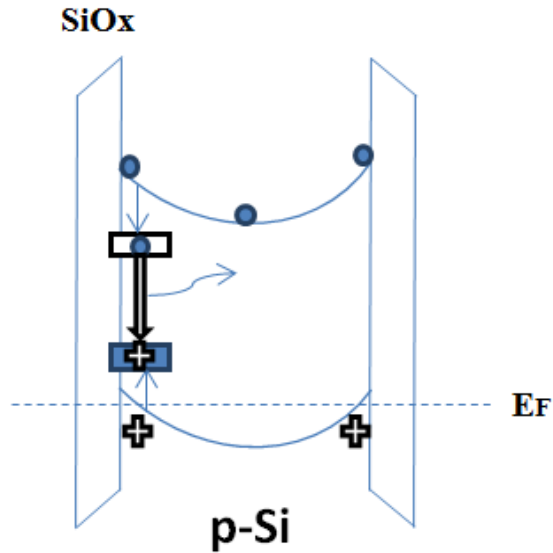

**Figure 4.** A sketch of the band structure indicating possible radiative recombination pathways through either interface states or defects associated with the dislocation line.

### 3.4 Devices

A thin Ti layer of 5 nm thick was evaporated, which was followed by a deposition of a Au layer of 20 nm through a shadow mask. Back side of the wafer was metalized by sputter deposited Aluminum layer of 100 nm. The electrical measurements were carried out to determine the device performance. All of the deuterium treated samples exhibited a strong rectifying characteristics. Au/D-Si/p-Si diode turned on at a forward bias of less than 0.4 V and had a reverse breakdown voltage in excess of 10 V. Figure 5 shows a typical the typical I-V characteristic of the heterojunction with the device structure illustrated in the insert of the figure. The device exhibits a series resistance of about 83 Ω at 1 V and 101 Ω at 0.5 V. The relatively low series resistances can be explained by the nature of the surface structure consisting of tips easing the carrier injection. The ideality factor $n = \frac{q}{kT}(\frac{\partial V}{\partial lnI})$ is around 3.9, which is relatively high as compared to bulk Si. The hig n value can be attributed to the presence of a disorder induced transport at the interface of Au/D-treated surface.

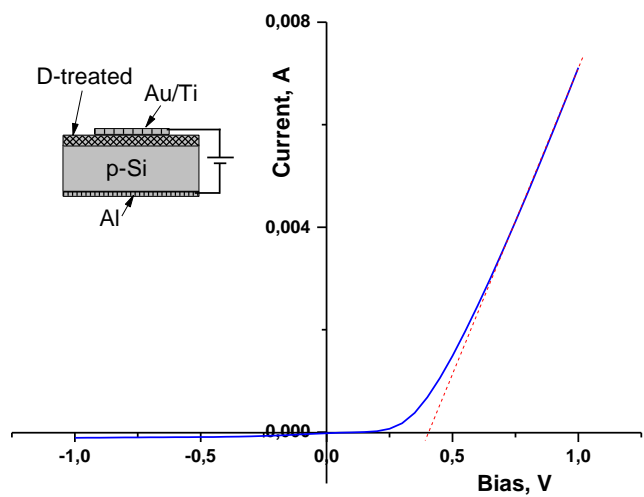

**Figure 5.** Typical current-voltage (I-V) characteristics of Au/Si devices as measured through Au contact. The current for all the structures shows rectifying behavior with a threshold voltage of 4.4V. N+-Si(111) 17h45'

## 4. CONCLUSION

Near infrared light emission of high quantum efficiency was obtained at telecommunication wavelengths at 1150nm, 1320 nm and 1550 nm from silicon treated by hydrogen and deuterium containing acid vapor. Each of these emissions can be reproducibly generated as a single peak in NIR region by controlling etching conditions. Such a modification of the surface of Si wafers results in a nanostructured highly porous surface exhibiting interesting optical properties such as a single luminescence peak at 1320 nm between 800 nm and 1800 nm. The emission overlaps with the D3 luminescence line suggesting probably a dislocation related origine for the peak. Due to the presence of oxygen and oxide encapsulation of nanostructured Si, carriers are likely efficiently trapped at the interface surface states thus lowering Schottky barriers in nanostructured Si/SiOx based devices. But also strong absorption in Si quantum structures enhances radiative recombination through either interface states or dislocation lines. Setting the peak position at 1550nm requires rougher surfaces thus indicating the role of defects. Further studies are required to determine whether disorder induced surface states or dislocations are responsible for the 1320 nm luminescence. The results of this work may lead to applications in Si based optoelectronics and photonics circuitries as well as optical interconnects for silicon photonics.

**Acknowledgement:** The author would like to thank P.Werner and V.Talalaev (MPI-Halle) for Si nanowire/luminescence support and U.Soderval (Chalmers Technology University) for MC2 nanofabrication access.